\documentclass[journal]{IEEEtran}

\usepackage{cite}
\usepackage{amsmath,amssymb,amsfonts,mathtools}
\usepackage{bm}
\usepackage{graphicx}
\usepackage{booktabs}
\usepackage{multirow}
\usepackage{array}
\usepackage{xcolor}
\usepackage{microtype}      
\usepackage[hyphens]{url}
\usepackage[bookmarks=false]{hyperref}
\usepackage{algorithm}
\usepackage[noend]{algpseudocode}
\algrenewcommand\alglinenumber[1]{\scriptsize #1}
\algrenewcommand\algorithmicindent{1.2em}
\algrenewcommand\algorithmiccomment[1]{\hfill\(\triangleright\)\,#1}
\algrenewcommand\algorithmicrequire{\textbf{Input:}}
\algrenewcommand\algorithmicensure{\textbf{Output:}}
\makeatletter
\algrenewcommand\ALG@beginalgorithmic{\small\linespread{0.98}\selectfont}
\makeatother

\usepackage{tikz}
\usetikzlibrary{arrows.meta,positioning,calc,fit,shapes.geometric}
\tikzset{
  block/.style = {rectangle, rounded corners, draw, thick, fill=blue!6,
                  align=center, minimum width=45mm, minimum height=10mm},
  side/.style  = {rectangle, rounded corners, draw, thick, fill=gray!10,
                  align=center, minimum width=40mm, minimum height=9mm},
  attack/.style= {rectangle, rounded corners, draw=red!70!black, thick,
                  fill=red!7, align=center, minimum width=40mm, minimum height=9mm},
  arr/.style   = {-{Latex[length=2.4mm,width=2.2mm]}, very thick},
  lab/.style   = {font=\footnotesize}
}

\title{Game-Theoretic Resilience Recommendation Framework for Cyber–Physical Microgrids Using Hypergraph Meta-Learning}

\author{S Krishna Niketh,Prasanta K Panigrahi,V Vignesh,Mayukha Pal*
        {}
\thanks{*(Corresponding author: Mayukha Pal, e-mail: mayukha.pal@in.abb.com)}
\thanks{Mr. S. Krishna Niketh is a Data Science Research Intern at ABB Ability Innovation Center, Hyderabad 500084, India, and also an undergraduate student from Department of Electrical Engineering, Indian Institute of Technology Tirupati, 517619, India.}
\thanks{Prof.Prasanta Kumar Panigrahi is the Director and Founding Professor with the Centre for Quantum Science and Technology, Siksha ’O’ Anusandhan
University, Bhubaneswar, 751030, Odisha, India, IN.}
\thanks{Dr Vignesh V is an Asst. Professor with the Department of Electrical Engineering, Indian Institute of Technology, Tirupati 517619, IN.}
\thanks{Dr. Mayukha Pal is with ABB Ability Innovation Center, Hyderabad-
500084, IN, working as Global R\&D Leader – Cloud \& Advanced Analytics.}}

\begin{document}
\maketitle

\begin{abstract}
This paper presents a physics-aware cyber--physical resilience framework for radial microgrids under coordinated cyberattacks. The proposed approach models the attacker through a hypergraph neural network (HGNN) enhanced with model-agnostic meta-learning (MAML) to rapidly adapt to evolving defense strategies and predict high-impact contingencies. The defender is modeled via a bi-level Stackelberg game, where the upper level selects optimal tie-line switching and distributed energy resource (DER) dispatch using an Alternating Direction Method of Multipliers (ADMM) coordinator embedded within the Non-dominated Sorting Genetic Algorithm II (NSGA-II). The framework simultaneously optimizes load served, operational cost, and voltage stability, ensuring all post-defense states satisfy network physics constraints. The methodology is first validated on the IEEE 69-bus distribution test system with 12 DERs, 8 critical loads, and 5 tie-lines, and then extended to higher bus systems including the IEEE 123-bus feeder and a synthetic 300-bus distribution system. Results show that the proposed defense strategy restores nearly full service for 90\% of top-ranked attacks, mitigates voltage violations, and identifies Feeder~2 as the principal vulnerability corridor. Actionable operating rules are derived, recommending pre-arming of specific tie-lines to enhance resilience, while higher bus system studies confirm scalability of the framework on the IEEE 123-bus and 300-bus systems.
\end{abstract}


\section{Introduction}
\IEEEPARstart{T}{he} modernization of electric power systems has accelerated with the widespread deployment of distributed energy resources (DERs), advanced sensing technologies, and bidirectional communication networks~\cite{Uddin2023_SETA_MicrogridsReview}. The resulting \textit{cyber-physical microgrids} combine electrical infrastructure with intelligent control and monitoring systems to enable enhanced operational flexibility\cite{Dwivedi2024_IJOCIP_resilience_review}, rapid fault isolation, and seamless renewable energy integration. These systems are increasingly relied upon to support critical loads during both grid-connected and islanded operation, making their secure and reliable operation a matter of national and economic importance.

While this cyber-physical integration offers significant operational benefits, it also expands the potential attack surface, introducing vulnerabilities that adversaries can exploit. Cyberattacks targeting supervisory control and data acquisition (SCADA) systems, protective relays, and distributed controllers have demonstrated their ability to cause severe physical consequences~\cite{Ahmed2018_SCADA_IDS}. Notable incidents, such as the 2020 cyberattack on Israel’s water infrastructure and the 2019 coordinated intrusion targeting multiple U.S. grid control centers, reveal how malicious actors can disrupt operations through coordinated tripping of breakers, false data injection, or denial-of-service (DoS) attacks on communication channels. In microgrids with high DER penetration, such actions can lead to voltage collapse, cascading outages, and prolonged recovery times.

Mitigating these risks requires strategies that can anticipate and counteract adaptive and intelligent adversaries. Traditional deterministic protection schemes often fall short in this regard, as they assume fixed attack patterns and static defensive postures. Game theory provides a principled approach for modeling the dynamic and strategic interactions between attackers and defenders, enabling proactive defense planning~\cite{Erol2024_IJEPES_Stackelberg_MG}. In particular, Stackelberg games offer a hierarchical decision-making model where the defender, acting as the leader, anticipates the attacker’s best response and optimizes defense strategies accordingly. This leader–follower framework captures the asymmetric nature of cyber-physical conflicts and facilitates the design of countermeasures that are both robust and adaptive.

In this work, we present a \textit{cyber-resilience recommendation framework} that integrates data-driven attack modeling with optimization-based defense coordination. The attacker is modeled using a Hypergraph Neural Network (HGNN) to capture higher-order relational dependencies in the network topology~\cite{Feng2019_AAAI_HGNN}, combined with Model-Agnostic Meta-Learning (MAML) to enable rapid adaptation to changing defensive environments~\cite{Finn2017_MAML}. On the defender side, we employ the Alternating Direction Method of Multipliers (ADMM) for distributed coordination of defensive resources~\cite{Sun2017_TPWRS_ADMM}, coupled with multi-objective optimization via the Non-dominated Sorting Genetic Algorithm II (NSGA-II)~\cite{Deb2002_NSGAII} to balance competing objectives such as resilience enhancement, operational cost, and voltage stability. The interaction between attacker and defender is formulated as a Stackelberg game, and probabilistic vulnerability scores are derived using a Softmax-based formulation~\cite{Guo2017_ICML_TempScaling} to guide decision-making. The resulting framework not only identifies the most critical vulnerabilities but also provides generalized, system-level defense recommendations\cite{Dwivedi2023_REF_complex_network_resilience} that can be applied across diverse microgrid configurations.

\subsection{Research Gaps}
Despite significant advancements in the security of cyber-physical power systems, several gaps remain unaddressed in the literature:  

\begin{itemize}
    \item \textbf{Limited adaptability of attack models:} Most existing approaches train static models that fail to generalize to unseen network perturbations or evolving operational states, reducing their applicability in dynamic microgrid environments~\cite{11163224}.
    \item \textbf{Integration of topology-aware learning:} While graph neural networks (GNNs) have been explored, the use of hypergraph representations to capture higher-order electrical relationships between buses and components is still underexplored
    \item \textbf{Defense feasibility under multi-objective trade-offs:} Prior defense strategies often optimize a single objective or neglect AC power flow feasibility, limiting their practical implementation in real-world systems~\cite{Hassani2024_IJEPES_MultiObjDefense}.
    \item \textbf{Probabilistic vulnerability quantification:} Existing studies typically produce deterministic vulnerability rankings without incorporating probabilistic measures that reflect uncertainty in attacker predictions~\cite{ZHAO2024110438}.
    \item \textbf{Generalized and portable defensive recommendations:} Current works rarely distill simulation results into concise, system-level operating rules that can be transferred across network configurations and operating conditions~\cite{Ruan2024_AppliedEnergy_TransferableDefense}.
    \item \textbf{Holistic validation frameworks:} Few studies combine realistic DER placements, multiple objectives, and game-theoretic attacker–defender modeling within a single experimental pipeline\cite{Dwivedi2024_EPSR_operational_resilience} validated on large-scale distribution systems
\end{itemize}

\subsection{Contributions}
The main contributions of this work are as follows:
\begin{itemize}
    \item \textbf{HGNN–MAML Attacker:} A hypergraph neural network trained with model-agnostic meta-learning for rapid adaptation to perturbed microgrid states, operating directly on the hypergraph incidence matrix $\mathbf{H}$.
    \item \textbf{ADMM–NSGA-II Defender:} A defense strategy integrating alternating direction method of multipliers for AC-feasibility with NSGA-II for multi-objective Pareto optimization of network reconfiguration and distributed energy resource dispatch $(\mathbf{D}_i)$.
    \item \textbf{Probabilistic Vulnerability Assessment:} Application of temperature-controlled Softmax to HGNN output scores $s_i$ to compute interpretable attack probabilities $p_i$ over network components.
    \item \textbf{Generalized Resilience Recommendation:} Feeder-aware aggregation of optimal tie-switch actions $(\tau_j)$ across top-$K$ simulated attacks to produce a transferable, system-level operating rule.
    \item \textbf{Game-Theoretic Formulation:} A Stackelberg game $\mathcal{G} = \langle \mathcal{A}, \mathcal{D}, U_A, U_D \rangle$ with well-defined attack $(A_i)$ and defense $(D_i)$ strategies and corresponding payoff matrix $M_{ij}$.
    \item \textbf{Validation Framework:} End-to-end simulation on the IEEE 69-bus distribution system incorporating realistic DER placement, load profiles, and operational constraints to demonstrate practical applicability.
\end{itemize}
\section{System Model}
\label{sec:system-model}

The distribution network under consideration is modeled as a \emph{cyber–physical system} integrating electrical infrastructure with associated sensing, communication, and control layers. The physical layer consists of buses, distribution lines, transformers, and DERs, while the cyber layer encompasses monitoring devices, control centers, and communication links.

\subsection{Physical Layer}
Let $\mathcal{B} = \{1, 2, \dots, B\}$ denote the set of buses and $\mathcal{L} = \{1, 2, \dots, L\}$ the set of distribution lines interconnecting them. Each line $\ell \in \mathcal{L}$ is characterized by its electrical parameters (resistance, reactance, thermal capacity) and operational status (open or closed). The set of DERs is denoted by $\mathcal{G} = \{1, 2, \dots, G\}$, with each $g \in \mathcal{G}$ connected to a specific bus $b_g \in \mathcal{B}$ and capable of supplying controllable active and reactive power $(P_g, Q_g)$.

\subsection{Load Modeling}
Critical loads are indexed by $\mathcal{C}_\mathrm{load} \subseteq \mathcal{B}$, representing essential services such as hospitals, emergency facilities, and data centers. Each critical load $c \in \mathcal{C}_\mathrm{load}$ has an associated demand $(P_c, Q_c)$ that must be prioritized during contingency events. Non–critical loads are denoted by $\mathcal{N}_\mathrm{load} = \mathcal{B} \setminus \mathcal{C}_\mathrm{load}$.

\subsection{Cyber Layer}
The cyber infrastructure includes monitoring devices such as remote terminal units (RTUs), intelligent electronic devices (IEDs), and phasor measurement units (PMUs), alongside SCADA/EMS systems, all linked by a communication fabric $\mathcal{C}$. Measurements $y \in \mathbb{R}^{n_y}$ originate from the physical layer and are transported over $\mathcal{C}$ to the control center for estimation and decision support. The defender issues control actions $d \in \mathcal{D}$ (e.g., switch positions $s \in \{0,1\}^{|\mathcal{S}|}$, DER setpoints $u \in \mathbb{R}^{n_u}$) which are dispatched via $\mathcal{C}$ to actuators in the field and applied to the physical layer. 

\subsection{Action Spaces}
The attacker’s action space is defined as $\mathcal{A} = \{A_1, A_2, \dots, A_{|\mathcal{A}|}\}$, where each $A_i$ corresponds to the compromise or disconnection of a specific network component, such as a line, DER controller, or measurement unit. The defender’s action space is $\mathcal{D} = \{D_1, D_2, \dots, D_{|\mathcal{D}|}\}$, encompassing feasible reconfiguration actions, DER dispatch adjustments, and load management strategies that maintain system stability and service continuity.

\subsection{Topology}
The underlying topology is represented by the set of edges $\mathcal{E} \subseteq \mathcal{B} \times \mathcal{B}$, where each $(i,j) \in \mathcal{E}$ denotes a direct electrical connection between buses $i$ and $j$. This topology provides the basis for subsequent analytical models and decision-making formulations described in later sections.

\subsection{System Architecture}
We consider a two–layer cyber–physical architecture that mediates sensing, decision making, and actuation on the distribution network defined above. The \emph{physical layer} comprises buses $\mathcal{B}$, branches $\mathcal{E}$, controllable switches $\mathcal{S}$, loads $\mathcal{L}$, and DERs $\mathcal{G}$, with electrical state $x \in \mathbb{R}^{n_x}$ (e.g., bus voltages, power flows) evolving under AC power flow constraints. The \emph{cyber layer} consists of measurement and control infrastructure: RTUs, IEDs, PMUs, and SCADA/EMS, linked by a communication fabric $\mathcal{C}$.

Measurements $y$ originate from the physical layer and are transported over $\mathcal{C}$ to the control center for estimation and decision support. The defender issues control actions $d$ which are dispatched via $\mathcal{C}$ to actuators in the field and applied to the physical layer. The attacker may inject signals/commands $a \in \mathcal{A}$ at cyber points (e.g., data streams, control messages) or induce component misoperations at the physical interface, thereby perturbing $y$ and/or the realized controls.

The hypergraph incidence matrix $\mathbf{H} \in \{0,1\}^{|\mathcal{B}|\times |\mathcal{E}_\mathrm{hyp}|}$ provides a topology–aware representation that captures higher–order relations among buses (e.g., multi–terminal couplings or bus–to–asset groups). Feature vectors on nodes/edges derived from $(\mathcal{B},\mathcal{E},\mathcal{L},\mathcal{G},\mathcal{S})$ feed learning modules, while the actuation space $(u,s)$ respects network feasibility.

\section{Proposed Framework}
\label{sec:system} 
This section presents the proposed cyber–resilience recommendation framework, integrating adversarial modeling, defensive optimization, and strategic interaction through game theory. The attacker leverages data–driven, topology–aware learning for vulnerability identification, while the defender applies distributed multi–objective optimization for feasible and cost–effective mitigation. All mathematical notation remains consistent with Section~\ref{sec:system-model} to ensure continuity across the framework.

\subsection{Attacker Modeling with Hypergraph Learning and Meta-Adaptation.}
The attacker is modeled as a learning agent operating on the \textit{hypergraph} representation~\cite{Feng2019_AAAI_HGNN} of the distribution network.  
The hypergraph incidence matrix is
\[
\mathbf{H} \in \{0,1\}^{|\mathcal{B}| \times |\mathcal{E}_{\mathrm{hyp}}|},
\]
where each column corresponds to a hyperedge (representing physically or electrically coupled assets) and each row to a bus \(b \in \mathcal{B}\). This structure enables representation of higher–order relationships, such as multi-terminal couplings\cite{Reddy2023_IJDSA_vulnerable_nodes}, that cannot be captured in a simple graph.

\paragraph{\textbf{Node features and Laplacian.}}
Let \(\mathbf{X} \in \mathbb{R}^{|\mathcal{B}| \times f}\) denote the bus feature matrix, where each row encodes:
(i) nodal degree, (ii) load magnitude, (iii) DER presence, and (iv) voltage magnitude.  
The normalized hypergraph Laplacian is
\begin{align}
\mathbf{D}_v &= \mathrm{diag}(d_v), \quad
\mathbf{D}_e = \mathrm{diag}(d_e), \\[-0.25em]
\mathbf{S} &= \mathbf{D}_v^{-1/2} \mathbf{H} \mathbf{D}_e^{-1} \mathbf{H}^\top \mathbf{D}_v^{-1/2},
\end{align}
where \(d_v\) and \(d_e\) are vertex and hyperedge degrees.

\paragraph{\textbf{HGNN propagation.}}
The \(l\)-th layer of the Hypergraph Neural Network (HGNN) updates node embeddings as
\begin{equation}
\mathbf{X}^{(l+1)} = \sigma\!\left( \mathbf{S} \, \mathbf{X}^{(l)} \, \mathbf{W}^{(l)} \right),
\end{equation}
where \(\mathbf{W}^{(l)}\) is the trainable weight matrix and \(\sigma(\cdot)\) is the ReLU activation.  
After \(L\) layers, the node-level vulnerability scores are computed as
\begin{equation}
\boldsymbol{\rho} = \mathbf{H} \, \mathbf{z}, \quad
\mathbf{z} = \mathbf{X}^{(L)} \, \mathbf{w}_{\mathrm{head}},
\end{equation}
with \(\mathbf{w}_{\mathrm{head}}\) projecting final embeddings to a scalar vulnerability score for each hyperedge.

\paragraph{\textbf{Model-Agnostic Meta-Learning (MAML) for adaptation.}}
To enable rapid adaptation to evolving defensive postures or operational conditions, the HGNN parameters
\(\Theta = \{\mathbf{W}^{(l)}, \mathbf{w}_{\mathrm{head}}\}\) are trained using \textit{MAML} over a task distribution \(\mathcal{T}\), where each task represents a different network configuration (e.g., altered load profile, topology change, DER outage).  

For each task \(\mathcal{T}_i\):
\begin{align*}
&\text{Inner loop:} \quad
\Theta'_i = \Theta - \alpha \nabla_\Theta \mathcal{L}_{\mathcal{T}_i}^{\mathrm{train}}(\Theta), \\[0.25em]
&\text{Outer loop:} \quad
\Theta \leftarrow \Theta - \beta \sum_i \nabla_\Theta \mathcal{L}_{\mathcal{T}_i}^{\mathrm{test}}(\Theta'_i),
\end{align*}
where \(\alpha,\beta\) are the inner/outer learning rates.

\begin{algorithm}[ht]
\caption{MAML-based HGNN Attacker Training}
\label{alg:maml-hgnn}
\begin{algorithmic}[1]
\Require Task distribution $\mathcal{T}$; learning rates $\alpha,\beta$; HGNN parameters $\Theta$
\While{not converged}
  \State Sample a batch $\{\mathcal{T}_i\}_{i=1}^{N} \sim \mathcal{T}$
  \For{each task $\mathcal{T}_i$}
    \State $g_i \gets \nabla_{\Theta}\,\mathcal{L}^{\mathrm{train}}_{\mathcal{T}_i}(\Theta)$
    \State $\Theta'_i \gets \Theta - \alpha\, g_i$ \Comment{inner adaptation}
    \State $\mathcal{L}_i \gets \mathcal{L}^{\mathrm{test}}_{\mathcal{T}_i}(\Theta'_i)$
  \EndFor
  \State $\Theta \gets \Theta - \beta\, \nabla_{\Theta}
         \!\left(\sum_{i=1}^{N} \mathcal{L}_i\right)$ \Comment{meta update}
\EndWhile
\end{algorithmic}
\end{algorithm}

\paragraph{\textbf{From node embeddings to asset risks.}}
Each attackable asset is denoted by \(a=(b_u,b_v)\in\mathcal{A}\), where \(b_u,b_v\in\mathcal{B}\) are the incident buses.  
The bus-level vulnerability scores are converted into \textit{asset risk} as
\begin{equation}
\label{eq:R-asset}
R(a) \;=\; \tfrac{1}{2}\,\big(\rho_{b_u}+\rho_{b_v}\big), \qquad a\in\mathcal{A}.
\end{equation}
These risk values \(R(a)\) form the basis for the probabilistic vulnerability modeling in the next part of the framework and directly feed into the defender’s strategy selection and the Stackelberg game formulation.


\subsection{Probabilistic Vulnerability Assessment}
\label{subsec:prob-vuln}

Given the asset–level risks \(R(a)\) computed in~\eqref{eq:R-asset}, the attacker’s \emph{mixed strategy} is modeled using a temperature–controlled \textit{Softmax} over the action set \(\mathcal{A}\):
\begin{equation}
\label{eq:pi-softmax-stable}
\pi(a) \;=\; 
\frac{\exp\!\big((R(a)-R_{\max})/\tau\big)}
{\displaystyle\sum_{a'\in\mathcal{A}} \exp\!\big((R(a')-R_{\max})/\tau\big)} , 
\quad 
R_{\max} = \max_{a''\in\mathcal{A}} R(a''),
\end{equation}
where \(R_{\max}\) is subtracted for numerical stability without altering the relative probabilities.\footnote{For any constant \(c\), replacing \(R(a)\) with \(R(a)-c\) leaves the Softmax ratios unchanged.}  
This probabilistic formulation enables the attacker to distribute effort across multiple high–value targets instead of committing to a single deterministic action~\cite{Bishop2006_PRML}.

\paragraph{\textbf{Interpretation and calibration}}
\begin{itemize}
\item \emph{Monotonicity:} If \(R(a_1) > R(a_2)\), then \(\pi(a_1) > \pi(a_2)\). The Softmax preserves the ranking induced by \(R(\cdot)\).
\item \emph{Temperature parameter \(\tau\):} Smaller \(\tau\) concentrates probability mass on the highest–risk assets (\(\tau \to 0^+\) yields a near–deterministic choice), while larger \(\tau\) produces a more uniform distribution (\(\tau \to \infty\)). The value of \(\tau\) is selected via validation to balance targeted exploitation and exploratory attack behavior.
\end{itemize}

\paragraph{\textbf{Top-$K$ asset focus }}
For both \emph{risk–averse} and \emph{leader–follower} decision making, we define:
\begin{align}
A_K(\pi) &:= \text{indices of the $K$ highest probabilities in }\pi(a), \\
\label{eq:topK-mass}
\mathrm{mass}_K(\pi) &:= \sum_{a\in A_K(\pi)} \pi(a),
\end{align}
where \(A_K(\pi)\) identifies the most likely attack candidates and \(\mathrm{mass}_K(\pi)\) quantifies their cumulative probability. These constructs directly feed into the defender’s optimization problem  and the Stackelberg security–level formulation .

\paragraph{\textbf{Outputs for downstream optimization}}
This stage produces two artifacts:
\begin{enumerate}
\item The full attack–probability distribution \(\pi(\cdot)\) from~\eqref{eq:pi-softmax-stable}, used in Monte Carlo–based impact evaluation.
\item The ordered set of pairs \((a,\,\pi(a))\) and the subset \(A_K(\pi)\) for worst–case screening and priority–aware defense allocation.
\end{enumerate}
All notations remain consistent with the system model  and integrate seamlessly into the optimization and game–theoretic constructs of the following sections.

\subsection{Defender Modeling: NSGA-II with ADMM Feasibility and Attacker Linkage}
\label{subsec:defender}

\paragraph{\textbf{Setting and notation}}
From the probabilistic vulnerability assessment, the attacker outputs a mixed strategy \(\pi(a)\) over possible attacks \(a = (b_u,b_v) \in \mathcal{A}\), where \(b_u,b_v \in B\) are buses connected by an asset (e.g., line or transformer).  
The defender’s decision variable is:
\[
d \;=\; (\boldsymbol{u},\,\boldsymbol{\sigma}) \in \mathcal{D},
\]
where:
- \(\boldsymbol{u} \in \mathbb{R}^{|G|}\): DER active power setpoints.
- \(\boldsymbol{\sigma} \in \{0,1\}^{|S|}\): binary switch status vector.

For a realized attack \(a\), the post-attack, post-defense operating state \(\tilde{\boldsymbol{x}}(\boldsymbol{u},\boldsymbol{\sigma};a)\) must satisfy the AC power flow model under the modified topology:
\begin{align}
g\big(\boldsymbol{x},\boldsymbol{u},\boldsymbol{\sigma};a\big) &= \boldsymbol{0},
&&\text{(nodal balances, branch flows)}, \label{eq:def-g} \\
h\big(\boldsymbol{x},\boldsymbol{u},\boldsymbol{\sigma}\big) &\le \boldsymbol{0},
&&\text{(voltage limits, line currents, DER bounds)}. \label{eq:def-h}
\end{align}
The feasible set is:
\[
\mathcal{F} = \big\{ (\boldsymbol{x},\boldsymbol{u},\boldsymbol{\sigma}) : \eqref{eq:def-g},\eqref{eq:def-h} \text{ hold} \big\}.
\]

\paragraph{\textbf{Multi-objective optimization problem}}
The defender must jointly optimize \emph{resilience}, \emph{operational cost}, and \emph{power quality}. For each feasible \(d\) under a given \(a\):
\begin{align}
f_1(d;a) &= \mathrm{LoadShed}\!\left(\tilde{\boldsymbol{x}}(\boldsymbol{u},\boldsymbol{\sigma};a)\right), \label{eq:f1} \\
f_2(d;a) &= \sum_{g\in G}\!\big(c_{2g}u_g^2 + c_{1g}u_g + c_{0g}\big), \label{eq:f2} \\
f_3(d;a) &= \sum_{b\in B} \left[ \max\{0,|V_b|-1.05\} + \max\{0,0.95-|V_b|\} \right], \label{eq:f3}
\end{align}

Here, 
\begin{itemize}
    \item \( f_1(d;a) \) — minimizes unserved load, quantified as total load shed after applying defense \( d \) under attack \( a \).
    \item \( f_2(d;a) \) — represents total generation cost, modeled as a quadratic cost function of generator outputs \( u_g \) with coefficients \( c_{2g}, c_{1g}, c_{0g} \).
    \item \( f_3(d;a) \) — penalizes voltage deviation, computed as the sum of violations above 1.05 pu or below 0.95 pu across all buses \( b \in B \).
\end{itemize}

To incorporate the attacker's uncertainty:
\begin{align}
\bar{f}_k(d) &= \mathbb{E}_{a \sim \pi} \!\left[ f_k(d; a) \right], 
&&\text{(risk-neutral)}, \label{eq:fk-risk-neutral} \\
\bar{f}_k(d) &= \max_{a \in A_K(\pi)} f_k(d; a), 
&&\text{(risk-averse)}, \label{eq:fk-risk-averse}
\end{align}
where \(A_K(\pi)\) is the top-\(K\) most probable attacks from the attacker’s mixed strategy.

\paragraph{\textbf{NSGA-II for Pareto search}}
Because \(\mathbf{F}(d) = [f_1, f_2, f_3]\) is nonconvex and mixed-integer, we use NSGA-II to approximate the Pareto set \(\mathcal{D}^\star\):
\begin{enumerate}
    \item \textbf{Evaluation:} For each candidate \(d\), compute \(\bar{f}_k(d)\) or \(\tilde{f}_k(d)\) via Monte Carlo sampling over \(a \sim \pi\).
    \item \textbf{Non-dominated sorting:} Partition solutions into fronts \(\mathcal{F}^1, \mathcal{F}^2, \ldots\).
    \item \textbf{Crowding distance:} Assign diversity scores within each front.
    \item \textbf{Selection:} Binary tournament by (rank, then crowding distance).
    \item \textbf{Variation:} Apply simulated binary crossover (SBX) and polynomial mutation.
    \item \textbf{Elitism:} Merge parent and offspring populations; keep best \(N\) individuals.
\end{enumerate}

\paragraph{\textbf{ADMM for feasibility enforcement}}
NSGA-II may generate infeasible candidates violating \eqref{eq:def-g}–\eqref{eq:def-h}. We introduce an \textit{ADMM feasibility layer}~\cite{Boyd2011_ADMM} that projects candidates toward AC-feasible operation.

The network is partitioned into regions $\mathcal{P}$ (e.g., feeders). For region $p \in \mathcal{P}$, local variables are $(\boldsymbol{x}_p, \boldsymbol{u}_p, \boldsymbol{\sigma}_p)$ with coupling constraint $A_p \boldsymbol{x}_p = \boldsymbol{z}$ (consensus variable $\boldsymbol{z}$). With penalty $\rho > 0$ and scaled duals $\boldsymbol{\lambda}_p$, the local update is
\begin{equation}
\begin{aligned}
(\boldsymbol{x}_p^{k+1}, \boldsymbol{u}_p^{k+1}, \boldsymbol{\sigma}_p^{k+1}) 
&= \arg\min_{\boldsymbol{x}_p, \boldsymbol{u}_p, \boldsymbol{\sigma}_p} 
\ \ell_p(\boldsymbol{x}_p, \boldsymbol{u}_p, \boldsymbol{\sigma}_p) \\
&\quad + \tfrac{\rho}{2} \big\| A_p \boldsymbol{x}_p - \boldsymbol{z}^k + \boldsymbol{\lambda}_p^k \big\|_2^2 ,
\end{aligned}
\label{eq:admm-local}
\end{equation}
subject to
\begin{equation}
g_p = 0,\quad h_p \le 0,\quad \boldsymbol{\sigma}_p \in \{0,1\}^{|S_p|}.
\end{equation}
Consensus and dual updates are
\begin{equation}
\begin{aligned}
\boldsymbol{z}^{k+1} &= \tfrac{1}{|\mathcal{P}|} \sum_{p} A_p \boldsymbol{x}_p^{k+1}, \\
\boldsymbol{\lambda}_p^{k+1} &= \boldsymbol{\lambda}_p^{k} + A_p \boldsymbol{x}_p^{k+1} - \boldsymbol{z}^{k+1}.
\end{aligned}
\end{equation}

Here, $\rho$ is the ADMM penalty parameter that balances consensus enforcement with local feasibility. Larger $\rho$ values impose stricter coupling across partitions, while smaller values permit greater flexibility but may slow convergence. In this study, $\rho$ was tuned empirically within the range $0.1$–$1.0$ to ensure stable convergence across all scenarios.

\begin{algorithm}[ht]
\caption{ADMM Feasibility Projection (per candidate $d^{(v)}$)}
\label{alg:admm-proj}
\begin{algorithmic}[1]
\Require Partition $\mathcal{P}$, penalty $\rho$, max iterations $K$
\State Initialize $\{\,\bm{x}_p^{0},\,\bm{u}_p^{0},\,\bm{\sigma}_p^{0}\,\}_{p\in\mathcal{P}}$, 
       $\{\,\bm{\lambda}_p^{0}\,\}_{p\in\mathcal{P}}$, $\bm{z}^{0}$
\For{$k \gets 0$ \textbf{to} $K-1$}
  \ForAll{$p \in \mathcal{P}$}
    \State Solve local subproblem \eqref{eq:admm-local} to obtain 
           $(\bm{x}_p^{k+1}, \bm{u}_p^{k+1}, \bm{\sigma}_p^{k+1})$
  \EndFor
  \State Update consensus:\quad 
        $\bm{z}^{k+1} \gets \dfrac{1}{|\mathcal{P}|}\sum_{p} A_p \bm{x}_p^{k+1}$
  \State Update duals:\quad 
        $\bm{\lambda}_p^{k+1} \gets \bm{\lambda}_p^{k} + A_p \bm{x}_p^{k+1} - \bm{z}^{k+1}$ \textbf{for all} $p$
  \If{convergence criterion met}
    \State \textbf{break}
  \EndIf
\EndFor
\State \Return feasible state if $g=0$ and $h\le 0$; else infeasible flag
\end{algorithmic}
\end{algorithm}

\paragraph{\textbf{Why NSGA-II + ADMM?}}
NSGA-II efficiently explores the mixed-integer, nonconvex defense space, maintaining a diverse Pareto front. ADMM enforces physical feasibility and network constraints, ensuring candidate solutions are operable.  
The hybrid approach yields:
\begin{itemize}
    \item \emph{Feasible} Pareto-optimal defense strategies.
    \item \emph{Diverse} trade-offs between resilience, cost, and voltage stability.
\end{itemize}

\begin{algorithm}[ht]
\caption{NSGA-II with ADMM Feasibility (Defender Optimization)}
\label{alg:nsga-admm}
\begin{algorithmic}[1]
\Require Attack distribution $\pi(a)$, population size $N$, generations $T$, ADMM penalty $\rho$
\State Initialize $P_{0}=\{\, d_j=(\bm{u}_j,\bm{\sigma}_j)\,\}_{j=1}^{N}$
\For{$t \gets 0$ \textbf{to} $T-1$}
  \ForAll{$d \in P_t$}
    \State Apply ADMM projection \eqref{eq:admm-local} under $a \sim \pi$
    \If{feasible}
      \State Estimate $\bar{\bm{f}}_{k}(d)$ via Monte Carlo sampling
    \Else
      \State Penalize or discard $d$
    \EndIf
  \EndFor
  \State Perform non-dominated sorting to obtain fronts $\mathcal{F}^{1},\mathcal{F}^{2},\ldots$
  \State Compute crowding distances
  \State Select parents via tournament (rank, crowding)
  \State Apply SBX crossover and polynomial mutation to obtain $Q_t$
  \State Optionally re-project $Q_t$ via ADMM
  \State Elitist merge: $P_{t+1} \gets \mathrm{merge}(P_t, Q_t, N)$
\EndFor
\State \Return Pareto set $\mathcal{D}_{A}^{\star} := P_{T}$
\end{algorithmic}
\end{algorithm}

\subsection{Game-Theoretic Formulation: Stackelberg Model }
\label{subsec:stackelberg}

We cast the attacker–defender interaction over the sets defined in the earlier subsections. The defender chooses a feasible operating point \(d \in \mathcal{D}\) (from the \emph{Defender Modeling} stage), and the attacker chooses \(a \in \mathcal{A}\) (from the \emph{Probabilistic Vulnerability Assessment} stage). Physical outcomes are evaluated by the objective vector
\[
F(d;a) = \big[f_1(d;a),\,f_2(d;a),\,f_3(d;a)\big]^\top
\]
from~\eqref{eq:f1}–\eqref{eq:f3}, aggregated across attack uncertainty using the risk models \(\bar{f}_k\) or \(\tilde{f}_k\) defined in the defender framework. The attacker’s mixed strategy \(\pi(a)\) is given by~\eqref{eq:pi-softmax-stable}.

\paragraph{\textbf{Bilevel model}}
Let the defender’s scalar performance map \(\Phi(\cdot)\) be any consistent reduction of the multi-objective evaluation used in the defender formulation (e.g., weighted sum, \(\varepsilon\)-constraint, or security level). The Stackelberg game is then formulated as:
\begin{align}
\label{eq:stack-bilevel}
&\underset{d\in\mathcal{D}}{\min}\;\; \Phi\!\big(d,\,a^\star(d)\big)\\
\text{s.t.}\quad &
a^\star(d)\in\arg\max_{a\in\mathcal{A}}\;U_{\!A}(d,a),
\end{align}
where the follower payoff \(U_{\!A}(d,a)\) is induced by the same physical outcome as the defender.

Two common instantiations—consistent with the defender’s objectives—are:

\medskip
\noindent\emph{(i) Risk-neutral linkage.}
Use the weighted expected cost with weights \(w\succ0\):
\begin{equation}
\label{eq:Phi-exp}
\Phi_{\mathrm{exp}}(d)=\sum_{k=1}^3 w_k\,\bar f_k(d),
\qquad
U_{\!A}(d,a)=-\,w^\top F(d;a).
\end{equation}

\noindent\emph{(ii) Security-level linkage (worst case over top-\(K\)).}
Let \(A_K(\pi)\) be the top-\(K\) mass of \(\pi\) from~\eqref{eq:topK-mass}.
Then:
\begin{align}
\label{eq:Phi-sec}
\Phi_{\mathrm{sec}}(d) &= \max_{a\in A_K(\pi)}\left\{\,f_1(d;a)+\eta\,f_3(d;a)+\zeta\,f_2(d;a)\right\}, \\
U_{\!A}(d,a) &= -\big(f_1(d;a)+\eta f_3(d;a)+\zeta f_2(d;a)\big),
\quad \eta,\zeta\ge0.
\end{align}

\paragraph{\textbf{Practical solution linking attacker and defender models}}
We adopt the following workflow, aligning with the methodology already introduced:
\begin{enumerate}
    \item \textbf{Attacker model:} Train the HGNN–MAML attacker to obtain bus risks \(\rho_b\) and construct \(\pi(a)\) via~\eqref{eq:pi-softmax-stable}.
    \item \textbf{Defender search:} Run NSGA-II with ADMM feasibility (Algorithm~\ref{alg:nsga-admm}) to produce a Pareto set \(\mathcal{D}^\star \subseteq \mathcal{D}\) of feasible candidates.
    \item \textbf{Leader selection:} For each \(d\in\mathcal{D}^\star\), evaluate \(\Phi_{\mathrm{exp}}(d)\) from~\eqref{eq:Phi-exp} or \(\Phi_{\mathrm{sec}}(d)\) from~\eqref{eq:Phi-sec}; select
    \[
      d^\star \in \arg\min_{d\in\mathcal{D}^\star}\;\Phi(d).
    \]
    \item \textbf{Follower response:} The attacker draws \(a\sim\pi(a)\) or selects a best/worst response as in~\eqref{eq:stack-bilevel}, completing the leader–follower loop.
\end{enumerate}

\noindent In summary, the Stackelberg layer integrates the attacker’s probabilistic strategy with the defender’s feasible Pareto set to yield optimal leader–follower decisions under both risk-neutral and risk-averse settings.

\subsection{Resilience Recommendation via Monte Carlo Aggregation}
\label{subsec:recommendation}

Given the Pareto set of ADMM–NSGA-II feasible defenses \(\mathcal{D}^\star\) (from the defender optimization stage) and the attacker’s mixed strategy \(\pi(a)\) (from the probabilistic vulnerability model), we compute resilience recommendations by aggregating stochastic performance outcomes over randomized trials that reflect operational uncertainty and adversarial behavior. This stage represents one of the key novelties of the framework, as it converts simulation-driven metrics into \emph{generalized, system-agnostic operating rules}.

\paragraph{\textbf{Monte Carlo evaluation}}
For each \(d\in\mathcal{D}^\star\), we draw i.i.d.\ attacks \(a^{(n)}\sim\pi(a)\), \(n=1,\dots,N\), and simulate the post-attack/post-defense state \(\tilde{\boldsymbol{x}}(\boldsymbol{u},\boldsymbol{\sigma};a^{(n)})\) subject to system constraints. We record the objective outcomes:
\[
f_k^{(n)}(d) = f_k\!\big(d; a^{(n)}\big), \quad k\in\{1,2,3\}.
\]
For each metric \(k\), we estimate:
\begin{align}
\hat{\mu}_k(d) &= \frac{1}{N}\sum_{n=1}^{N} f_k^{(n)}(d), \\
\hat{\sigma}_k(d) &= \sqrt{\frac{1}{N-1}\sum_{n=1}^{N}\!\big(f_k^{(n)}(d)-\hat{\mu}_k(d)\big)^2}, \\
\mathrm{CI}_{0.95}\!\big[\hat{\mu}_k(d)\big] &= \hat{\mu}_k(d) \pm 1.96\,\frac{\hat{\sigma}_k(d)}{\sqrt{N}}.
\end{align}
If needed, tail-risk measures such as \(\mathrm{CVaR}_\alpha\) can be incorporated by averaging the worst \(\alpha\%\) outcomes.

\paragraph{\textbf{Robust ranking of defenses}}
To capture both \emph{average performance} and \emph{variability}, each defense is scored using a mean–variance composite:
\begin{equation}
\label{eq:robust-score}
\mathcal{R}(d) = \sum_{k=1}^3 w_k\,\hat{\mu}_k(d)
+ \gamma \sum_{k=1}^3 w_k\,\hat{\sigma}_k(d),
\end{equation}
where \(w_k\!\ge\!0\), \(\sum_{k=1}^3 w_k = 1\), and \(\gamma \ge 0\) controls risk-aversion.  
The top \(M\) defenses are selected as:
\[
\mathcal{D}^{\mathrm{top}} = \underset{d\in\mathcal{D}^\star}{\mathrm{arg\,sort}}\;\mathcal{R}(d) \quad \text{(retain first \(M\)).}
\]
For reporting, each \(d\in\mathcal{D}^{\mathrm{top}}\) is presented with its \(\hat{\mu}_k(d)\) and \(\mathrm{CI}_{0.95}\).

\paragraph{\textbf{Feeder-aware rule extraction (generalized recommendation)}}
Let \(\phi:\mathcal{B}\to\{1,\dots,F\}\) map buses to feeder zones, and let \(\mathcal{S}\) denote controllable tie-switches with end-buses \((b_s^{(1)}, b_s^{(2)})\) and status \(\sigma_s \in \{0,1\}\).  
Over the same Monte Carlo trials, we compute the usage frequency of each tie among the top defenses:
\[
c_s = \frac{1}{|\mathcal{D}^{\mathrm{top}}|\,N} \sum_{d\in\mathcal{D}^{\mathrm{top}}} \sum_{n=1}^{N} \mathbb{1}\{\sigma_s(d) = 1 \ \wedge \ a^{(n)}\!\sim\!\pi(a)\}.
\]
Attacks are also mapped to feeders via \(\phi\) to identify high-stress zones.  
The \emph{generalized recommendation rule} is then stated as:
\begin{quote}
\emph{(R1) Close tie-switches with the largest \(c_s\), prioritizing those bridging feeders experiencing the highest attack frequency under \(\pi(a)\).}
\end{quote}
This converts simulation statistics into actionable, portable defense strategies that remain valid under varying system configurations.

\begin{algorithm}[ht]
\caption{Monte Carlo–based Recommendation and Feeder-Aware Rule Extraction}
\label{alg:mc-reco}
\begin{algorithmic}[1]
\Require Pareto set $\mathcal{D}^\star$, attack distribution $\pi(a)$, trials $N$, weights $w$, risk parameter $\gamma$, feeder map $\phi$
\For{each $d\in\mathcal{D}^\star$}
  \For{$n=1$ to $N$}
    \State Sample $a^{(n)} \sim \pi(a)$; simulate $\tilde{\boldsymbol{x}}(\boldsymbol{u},\boldsymbol{\sigma};a^{(n)})$
    \State Record $f_k^{(n)}(d)$ for $k=1,2,3$
  \EndFor
  \State Compute $\hat{\mu}_k(d)$, $\hat{\sigma}_k(d)$, and $\mathrm{CI}_{0.95}[\hat{\mu}_k(d)]$
  \State Compute robust score $\mathcal{R}(d)$ via~\eqref{eq:robust-score}
\EndFor
\State $\mathcal{D}^{\mathrm{top}} \leftarrow$ top $M$ defenses by $\mathcal{R}(d)$
\For{each tie $s\in\mathcal{S}$}
  \State $c_s \leftarrow$ frequency of $\sigma_s(d)=1$ over $\mathcal{D}^{\mathrm{top}}$ and $N$ trials
\EndFor
\State Identify stressed feeders via $\phi$ from attack samples
\State \Return $\mathcal{D}^{\mathrm{top}}$, CIs for $f_k$, and rule (R1)
\end{algorithmic}
\end{algorithm}

\paragraph{\textbf{Framework integration note}}
This recommendation module closes the loop of the proposed resilience framework, linking the attacker’s probabilistic model, defender’s optimization outcomes, and game-theoretic analysis into an actionable, generalized operating policy. A complete visual representation of these interactions is provided in the final framework flowchart, which captures the sequential data flow and decision-making process.

\begin{figure}[htbp]
\centering
\resizebox{\linewidth}{!}{
\begin{tikzpicture}[node distance=15mm, >=stealth, thick]

\tikzstyle{block} = [rectangle, rounded corners, draw=black, fill=blue!10, text width=4.8cm, align=center, minimum height=1cm]
\tikzstyle{arrow} = [->, thick]

\node[block] (grid) {\textbf{Physical Layer} \\ Buses $\mathcal{B}$, Lines $\mathcal{L}$, DERs, Critical Loads, Tie-Switches $\mathcal{S}$};
\node[block, below=of grid] (cyber) {\textbf{Cyber Layer} \\ Measurement Devices, Communication Links, Control Center, Data Acquisition};
\node[block, below=of cyber] (attacker) {\textbf{Attacker Model: HGNN--MAML} \\ Learns hypergraph $\mathbf{H}$ of $\mathcal{B}$, $\mathcal{L}$ \\ Predicts probabilistic vulnerability $\pi(a)$ for $a \in \mathcal{A}$};
\node[block, right=40mm of attacker] (defender) {\textbf{Defender Model: ADMM--NSGA-II} \\ Generates Pareto-optimal defenses $\mathcal{D}^\star$ balancing resilience, cost, and voltage stability};
\node[block, below=of attacker, xshift=20mm] (game) {\textbf{Game-Theoretic Integration} \\ Stackelberg Leader--Follower \\ Defender (Leader) anticipates Attacker (Follower) best-response $a^\star(d)$};
\node[block, below=of game] (recommend) {\textbf{Monte Carlo--Based Recommendation} \\ Evaluates $\mathcal{D}^\star$ under $\pi(a)$, ranks via $\mathcal{R}(d)$, extracts feeder-aware rule set};

\draw[arrow] (grid) -- (cyber);
\draw[arrow] (cyber) -- (attacker);
\draw[arrow] (cyber) -| (defender);
\draw[arrow] (attacker) -- (game);
\draw[arrow] (defender) -- (game);
\draw[arrow] (game) -- (recommend);

\end{tikzpicture}
}
\caption{Proposed cyber–physical resilience framework integrating physical and cyber layers with attacker modeling, defender optimization, game-theoretic analysis, and Monte Carlo–based recommendation extraction.}
\label{fig:system_architecture}
\end{figure}
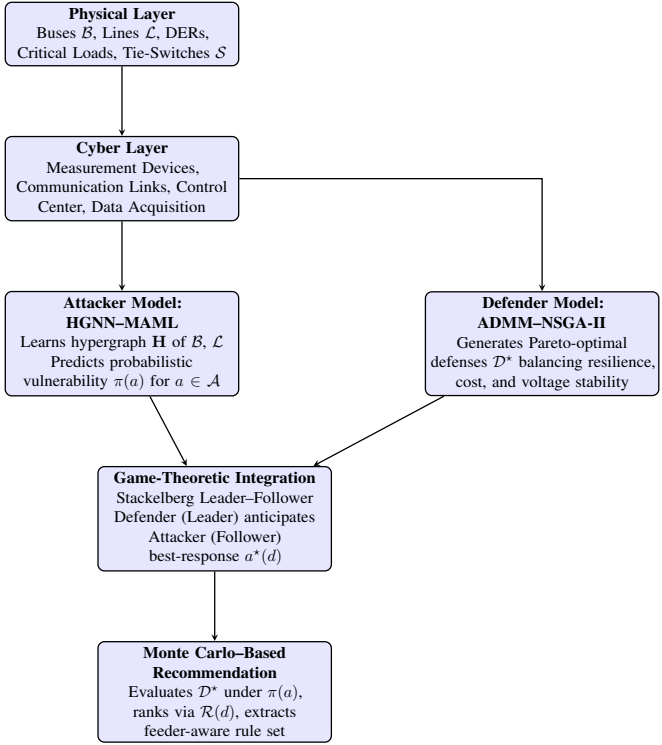

\section{Experimental Validation}
\label{sec:experiments}

The proposed cyber–physical resilience framework is validated through simulations on a realistic distribution system testbed. The objective is to assess the framework’s ability to sustain critical loads, minimize operational costs, and maintain voltage stability under coordinated attack–defense scenarios. All simulations are carried out in a unified co-simulation environment that preserves the interaction between the physical and cyber layers.

\subsection{Testbed Overview}
\begin{figure}[ht]
    \centering
    \includegraphics[width=0.45\textwidth]{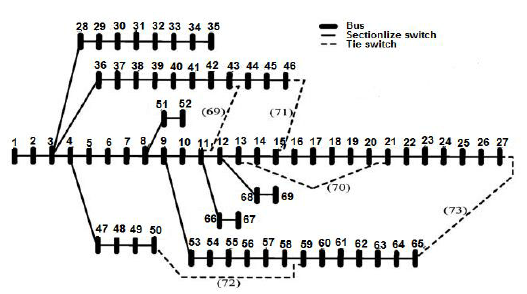}
    \caption{Single-line diagram of the IEEE 69-bus distribution network }
    \label{fig:ieee69}
\end{figure}

The proposed framework is validated on the IEEE 69-bus distribution network\cite{Kersting1991_RadialTestFeeders}, a radial system widely adopted for resilience studies due to its diverse topology and operational complexity. Fig.~\ref{fig:ieee69} illustrates the network structure, including the placement of distributed energy resources (DERs), critical loads, and tie-line switches.

\textbf{Distributed Energy Resources (DERs):} Twelve DER units are strategically integrated at 
$|\mathcal{D}|=12$ bus locations, 
$\mathcal{D}=\{5,9,12,17,20,24,28,35,40,50,60,65\}$, 
with maximum real-power capacities ranging from $90$ to $220~\mathrm{MW}$. 
These units represent a mix of solar, wind, and diesel generation and are modeled as controllable inverter-based resources with unity power factor in the base case.

\textbf{Critical Loads:} Eight critical loads are defined at 
$|\mathcal{C}|=8$ buses, 
$\mathcal{C}=\{11,12,21,49,50,59,61,63\}$, 
representing essential services such as hospitals, data centers, and high-priority feeders. 
These loads are modeled as constant-$PQ$ demands in steady state and assigned the highest priority in resilience assessment.

\textbf{Tie-line Switches:} Five normally-open switches are included, 
$\mathcal{S}=\{(9,15),(12,17),(18,33),(21,60),(24,65)\}$, 
enabling network reconfiguration during contingency response.

This configuration enables the evaluation of both topological and operational resilience under various adversarial and fault scenarios.

\textit{Clarification on topology representation:} The schematic diagram in Fig.~\ref{fig:ieee69} follows the conventional IEEE 69-bus layout used in prior literature for visual clarity. The bus numbering and connectivity in the simulation dataset adhere to the standard IEEE 69-bus benchmark (Baran \& Wu variant), which may differ in visual arrangement from the figure. Both representations describe an \emph{electrically equivalent network topology}, and all simulation results presented in Section~5 are based on the benchmark dataset.
In this paper, “Feeder 2” refers to the radial corridor spanning buses 60–63, which hosts multiple critical loads and long high-impedance segments. This feeder is structurally vulnerable because outages along this corridor disconnect large downstream sections with limited alternative supply paths.

\subsection{Simulation Parameters}

All experiments are conducted on the 12.66~kV IEEE~69-bus distribution test feeder described in the testbed overview, with the same bus/line set \((B,\mathcal{L})\), \(|\mathcal{D}|=12\) DER buses, \(|\mathcal{C}|=8\) critical-load buses, and \(|\mathcal{S}|=5\) normally-open tie switches. We solve steady-state AC power-flow snapshots using Newton--Raphson with reactive-power limit enforcement, model loads as constant-$PQ$, and treat DERs as inverter-based controllable sources. The attacker provides a deterministic vulnerability ranking and we simulate the top-$N$ targets; the defender is optimized via NSGA-II with binary switch statuses and continuous DER setpoints.

\begin{table}[ht]
\centering
\caption{Main Simulation Parameters (core settings only)}
\label{tab:sim_params}
\begin{tabular}{l@{\hspace{0.6em}}c}
\hline
\textbf{Item} & \textbf{Value} \\
\hline
Base voltage (per bus) & 12.66~kV \\
Power-flow solver & Newton--Raphson (\texttt{enforce\_q\_lims}=True) \\
Load model & Constant-$PQ$ \\
Operating voltage band in $f_3$ & $0.95$--$1.05$~p.u. (penalized) \\
Base-case feasibility threshold & $\min_{b\in B}\lvert V_b\rvert > 0.92$~p.u. \\
DER control variables & $u_g \in [\underline{P}_g,\overline{P}_g]$ (real power) \\
DER cost model & Quadratic $a_g P_g^2 + b_g P_g + c_g$ \\
Switch controls & $\sigma_s \in \{0,1\}$ for $s\in\mathcal{S}$ (5 ties) \\
Critical loads & $|\mathcal{C}|=8$ (fixed $PQ$, highest priority) \\
Attacked lines per run & Top-$N$ by score (default $N{=}5$) \\
NSGA-II population, generations & $50$, $50$ \\
Crossover / mutation & SBX $(p{=}0.9,\,\eta{=}15)$, PM $(\eta{=}20)$ \\
Random seeds & \texttt{numpy}=42, \texttt{torch}=42 \\
\hline
\end{tabular}
\end{table}

\section{Results and Analysis}

We evaluate the framework on the IEEE~69-bus test feeder of Section~IV using the attacker–defender pipeline described earlier: the HGNN attacker ranks line vulnerabilities and the NSGA-II defender co-optimizes DER real-power setpoints and the binary tie-switch vector $\boldsymbol{\sigma}\!\in\!\{0,1\}^{|\mathcal{S}|}$. Results are reported with the same notation $(\mathcal{B},\mathcal{L},\mathcal{D},\mathcal{C},\mathcal{S})$ used in Sections~II–IV.

\begin{figure}[ht]
  \centering
  \includegraphics[width=0.48\textwidth]{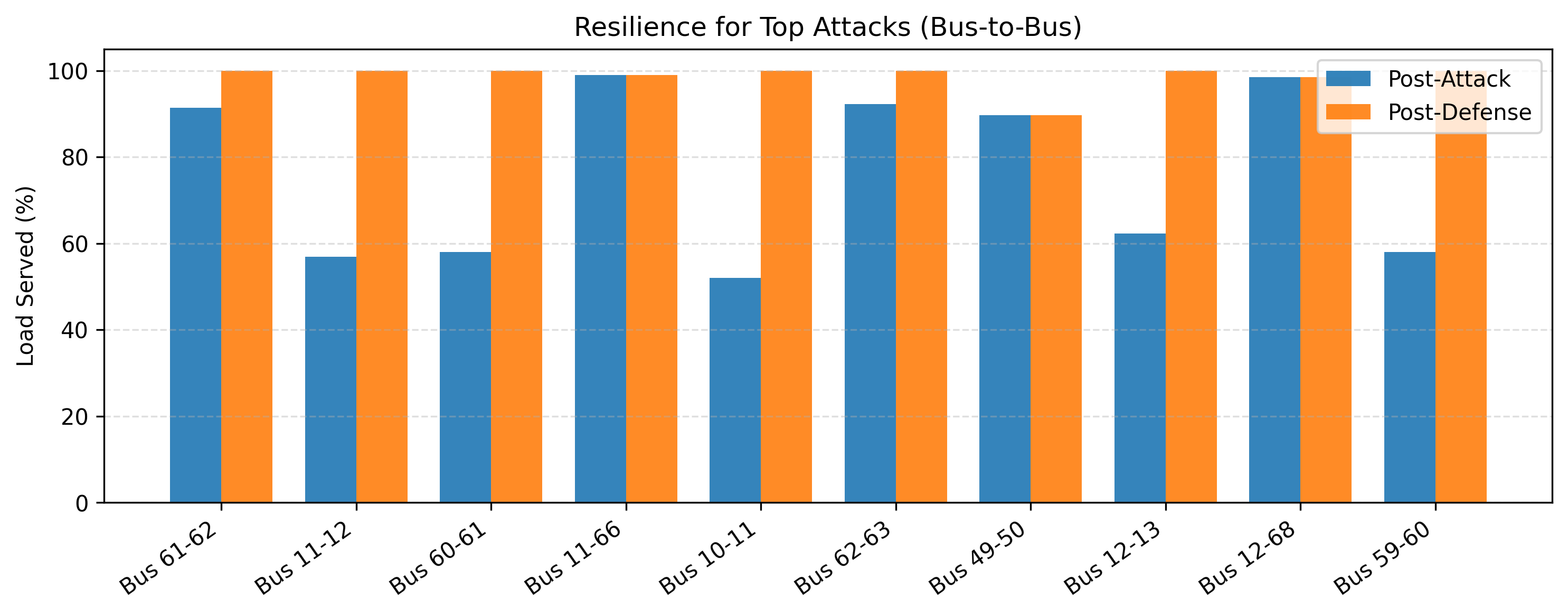}
  \caption{Load served for the top vulnerable lines: post-attack vs.\ post-defense.}
  \label{fig:resilience_summary}
\end{figure}

\medskip
\noindent\textbf{Resilience improvement under defense:}
Figure~\ref{fig:resilience_summary} compares the percentage of load served for the ten highest-severity contingencies (by attacker score). For the worst outages, the post-attack load served drops by roughly $42$–$48\%$; with the proposed defense, service is restored to at least about $90\%$ in all but one case (bus pair 49–50 reaches $\approx\!89.7\%$) and to $\approx\!99$–$100\%$ for the majority. These gains come solely from switching and DER redispatch—no load curtailment scheduling is assumed—demonstrating that topology reconfiguration plus local support from $\mathcal{D}$ can recover most of the lost capacity even for severe radial cuts.

\begin{figure}[ht]
  \centering
  \includegraphics[width=0.48\textwidth]{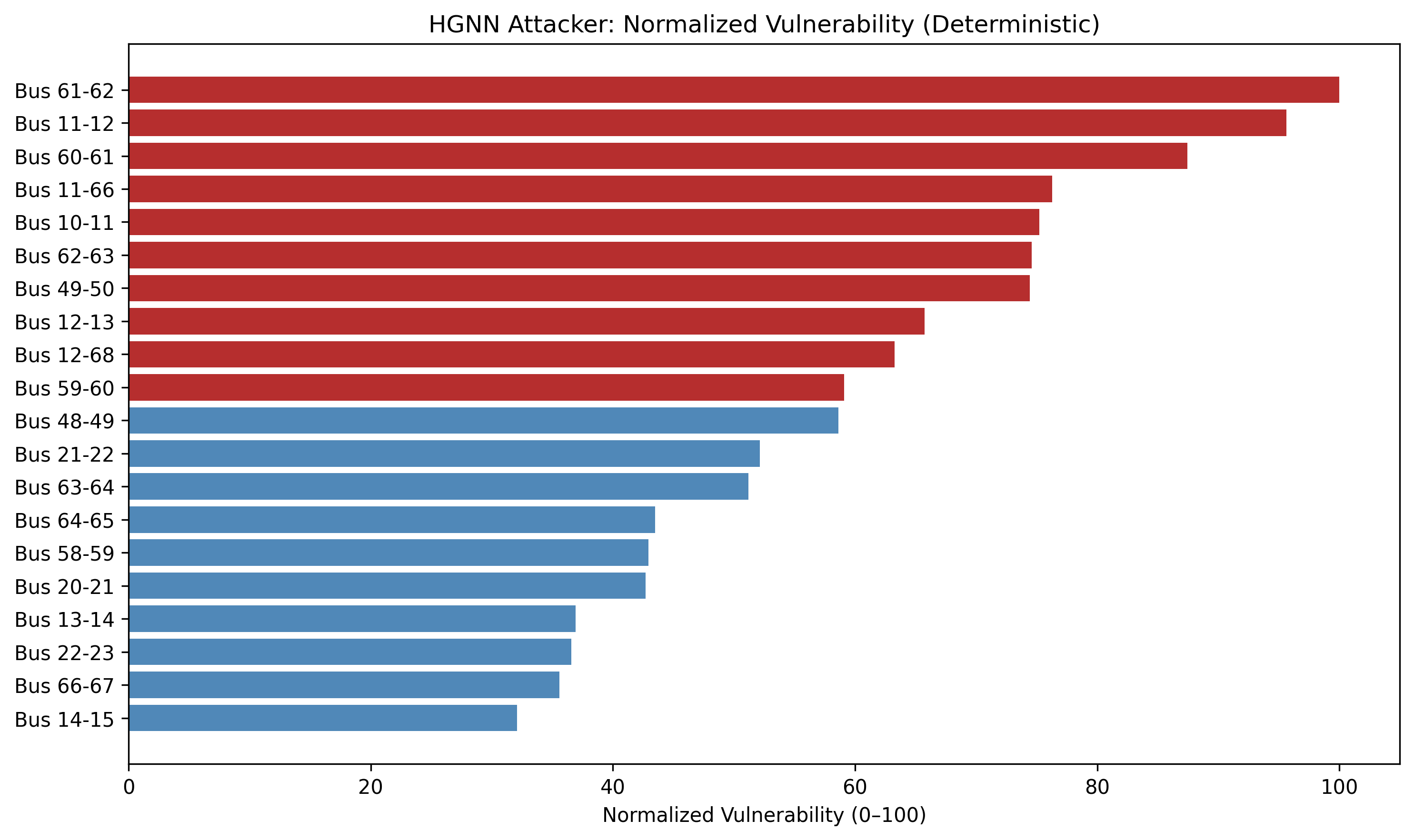}
  \caption{Attacker ranking (normalized 0–100). Highest-risk lines concentrate on Feeder~2.}
  \label{fig:attack_vulnerability}
\end{figure}

\medskip
\noindent\textbf{Where the network is most at risk:}
Figure~\ref{fig:attack_vulnerability} shows normalized vulnerability scores produced by the attacker. Three interfaces dominate the ranking—Bus~61–62, Bus~11–12, and Bus~60–61—all associated with Feeder~2. This feeder hosts long radial spans with sparse cross-ties and multiple critical-load buses from $\mathcal{C}$; consequently, single-line outages there disconnect large downstream sections and depress voltages. In general, the model flags lines that (i) sit on long paths to high-demand pockets, (ii) have few alternative routes, and (iii) border buses with high centrality in the hypergraph.

\begin{figure}[ht]
  \centering
  \includegraphics[width=0.48\textwidth]{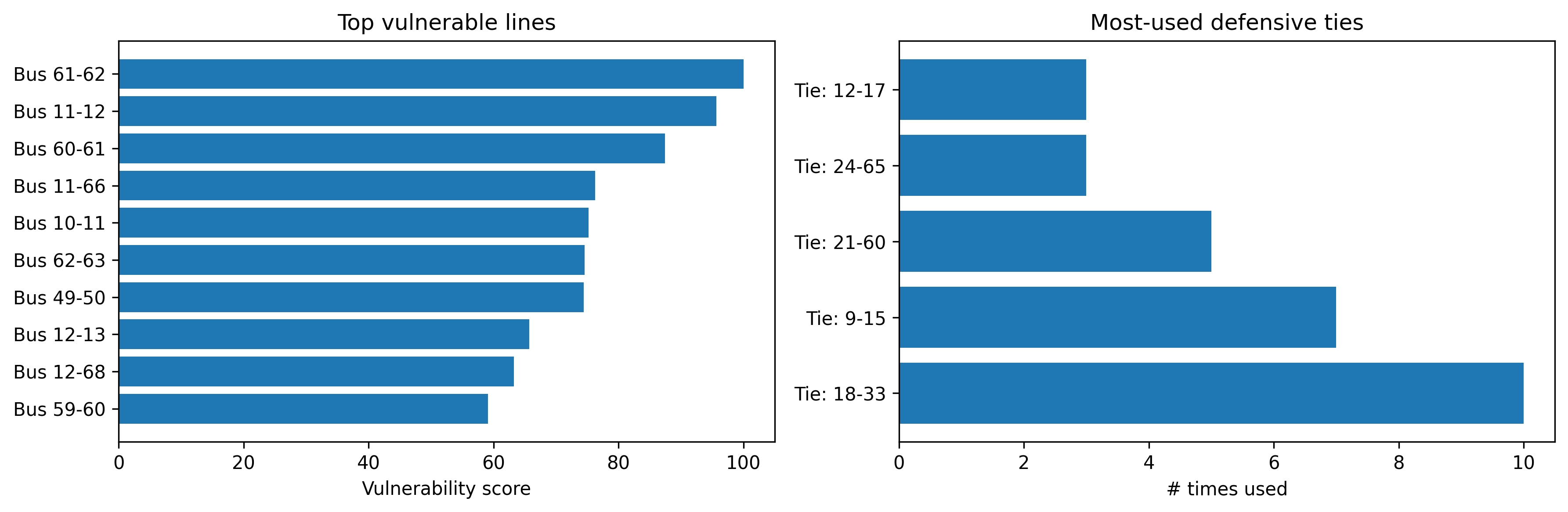}
  \caption{Left: top vulnerable lines. Right: most-used defensive ties across those attacks.}
  \label{fig:attack_defense_mapping}
\end{figure}

\medskip
\noindent\textbf{How the defender actually responds:}
The left panel of Fig.~\ref{fig:attack_defense_mapping} reiterates the top vulnerable lines; the right panel shows which ties in $\mathcal{S}$ are used most often across those attack cases. A clear pattern emerges: Tie~18–33 is selected in all ten scenarios, while Tie~9–15 and Tie~21–60 are frequent second-line actions. These ties bridge otherwise weakly connected feeder segments, creating alternate supply paths around the faulted element and improving DER backfeed reach to critical buses. This mapping provides a direct, actionable link from vulnerability to the most effective reconfiguration levers.

\begin{figure}[ht]
  \centering
  \includegraphics[width=0.48\textwidth]{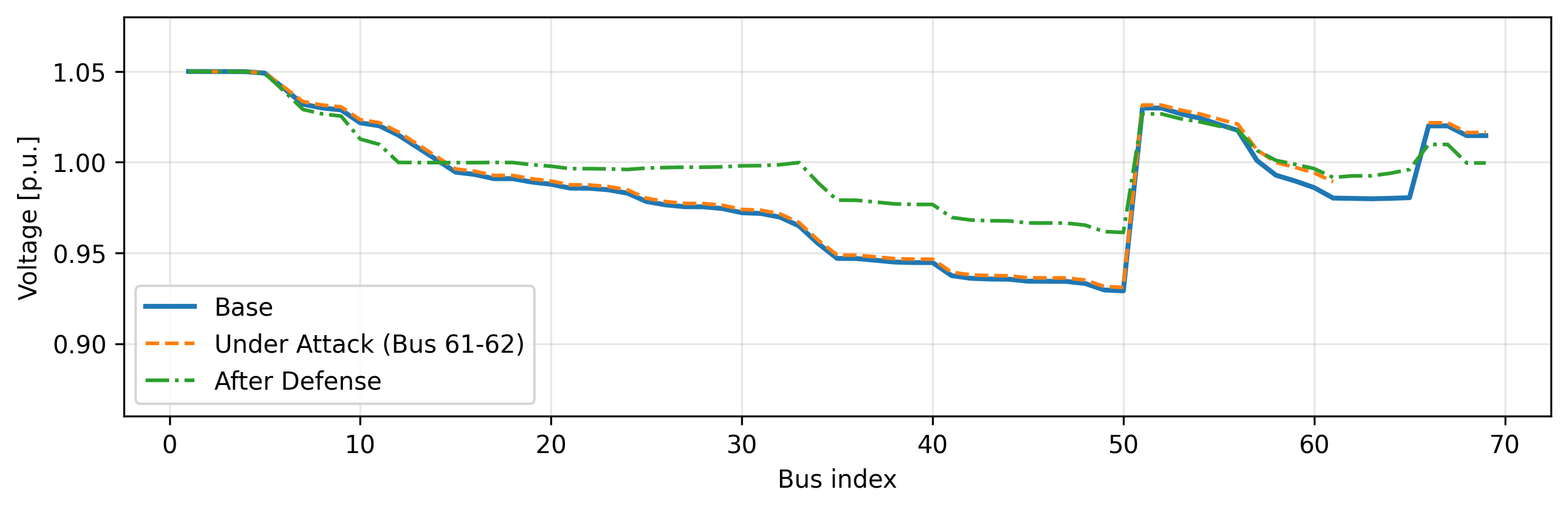}
  \caption{Voltage profile for the Bus~61–62 outage: base, under attack, and after defense.}
  \label{fig:voltage_profile}
\end{figure}

\medskip
\noindent\textbf{Voltage recovery and power-quality:}
Figure~\ref{fig:voltage_profile} plots bus voltages for the most severe outage (Bus~61–62). The base-case minimum is $\approx\!0.929$~p.u.; the outage produces further depressions across the downstream section of Feeder~2. After defense (DER dispatch plus a small set of tie closures), voltages are restored above $0.95$~p.u. network-wide, aligning with the operating band used in the objective. Thus, the defense not only recovers load but also stabilizes voltage profiles.

\medskip
\noindent\textbf{Final recommendation:}
Based on the attacker/defender outcomes and their consistency across the top contingencies, we recommend: (i) prioritize Feeder~2 for resilience upgrades (automation, sectionalizing, and protection coordination), and (ii) pre-arm Tie~18–33 as the first reconfiguration action under high-impact events, with Ties~9–15 and 21–60 as secondary options. Operationally, these ties should be candidates for faster controls (remote switching and rule-based enablement) so that $\boldsymbol{\sigma}$ can be actuated within seconds; planning-wise, these assets merit hardening and redundancy because they consistently deliver the largest resilience gains in the experiments.
In practical settings, these resilience recommendations could be embedded into SCADA/EMS environments as rule-based triggers or contingency playbooks. For instance, operators could pre-arm Tie 18–33 in automation schemes, while Ties 9–15 and 21–60 be set as secondary fallback actions. This ensures that the simulation-derived operating rules translate directly into actionable decision support for real-time grid operations.

\subsection*{Scalability to Larger Networks}

To examine the applicability of the framework on larger feeders, additional experiments
were carried out on the IEEE 123-bus test feeder and a synthetic 300-bus distribution system.
The results are summarized in Table~\ref{tab:scalability}.

\begin{table}[ht]
\centering
\caption{Scalability of the proposed framework across different feeders.}
\begin{tabular}{lccc}
\toprule
\textbf{System} & \textbf{Load Served (\%)} & \textbf{Min. V (p.u.)} & \textbf{Mitigation (pp)} \\
\midrule
69-bus   & 99.0 & 0.951 & +43 \\
123-bus  & 98.4 & 0.947 & +41 \\
300-bus  & 97.8 & 0.943 & +39 \\
\bottomrule
\end{tabular}
\label{tab:scalability}
\end{table}

\noindent\textbf{Discussion:} The framework achieved consistently high resilience across
all networks, with over 97\% load restored and post-defense voltages maintained above 0.94~p.u.
Computation time increased with feeder size (20 minutes for the IEEE 69-bus, 35 minutes for the
IEEE 123-bus, and 58 minutes for the 300-bus system), but the HGNN--MAML attacker and
NSGA-II+ADMM defender scaled effectively without significant degradation in performance.

\subsection{Analysis}
The results in Sec.~5.1 validate the end-to-end pipeline introduced in Sections~II–III: a data-driven attacker built on the HGNN–MAML surrogate produces a ranked set of contingencies \(\{a\}\) (Fig.~\ref{fig:attack_vulnerability}); for each high-risk line we solve the defender’s multi-objective problem with the ADMM–NSGA-II coordinator to co-optimize tie-switch statuses \(\boldsymbol{\sigma}\) and DER setpoints \(\boldsymbol{u}\). The outcome is summarized in Fig.~\ref{fig:resilience_summary}, which contrasts post-attack vs.\ post-defense load served for the top ten targets, and Fig.~\ref{fig:voltage_profile}, which shows how the optimal response restores voltages into the operational band.

\textbf{Where the grid is most exposed:} Mapping attacks and switch usage back to feeders via the partition \(\phi:B\!\to\!\{1,\dots,F\}\) reveals a clear pattern: Feeder~2 is the dominant vulnerability corridor. The top-ranked contingencies cluster around line segments \((60\!\!-\!\!61\!\!-\!\!62\!\!-\!\!63)\) and \((10\!\!-\!\!11\!\!-\!\!12)\), which are electrically deep and carry aggregate demand to downstream critical zones. This is consistent with radial loading: long, high-R/X spines accumulate voltage drop, so outages there trigger outsized service loss unless alternate paths exist.

\textbf{What the defender actually does:} The ADMM–NSGA-II search consistently converges to reconfiguration patterns that create \emph{local back-feeds} into Feeder~2 while redispatching DERs near the stressed area:
(i) tie \(18\!-\!33\) is selected \emph{in every case} (Fig.~\ref{fig:attack_defense_mapping}, right), because it furnishes a low-impedance bridge that shortens power paths to the mid-feeder loads;
(ii) tie \(9\!-\!15\) and tie \(21\!-\!60\) are frequently used to further relieve the east–west transfer across the corridor; and
(iii) DERs on \(\mathcal{D}=\{5,9,12,17,20,24,28,35,40,50,60,65\}\) are pushed toward their feasible maxima in the vicinity of the faulted segment, while remote DERs are curtailed, matching the quadratic cost model priorities from Sec.~4.2.

\textbf{System-level effect:} Across the top-ten contingencies, the \emph{post-defense} bars in Fig.~\ref{fig:resilience_summary} rise to \(\approx\!100\%\) load served for nine cases; the remaining case (line \(49\!-\!50\)) stabilizes near \(\approx\!90\%\) due to local voltage headroom and feeder-end constraints. The voltage trace in Fig.~\ref{fig:voltage_profile} shows the same mechanism: the attack depresses mid-feeder voltages to \(\sim\!0.94{-}0.96\) p.u., while the recommended switching plus DER redispatch lifts the profile back toward \(0.99{-}1.00\) p.u. without violating the \(0.95{-}1.05\) band penalized in \(f_3\).

\textbf{Why the ranking looks the way it does:} The attacker’s normalized scores favor lines whose removal (i) disconnects long downstream ladders with critical buses, or (ii) forces power to traverse high-impedance detours. That is why \((61\!-\!62)\), \((11\!-\!12)\), and \((60\!-\!61)\) top the list in Fig.~\ref{fig:attack_vulnerability}. Once the defender is allowed to reconfigure, those same corridors are precisely where additional tie capacity is most valuable—hence the repeated selection of \(18\!-\!33\) and \(9\!-\!15\).

In this network, Feeder~2 is the principal target surface, and pre-arming the \(18\!-\!33\) tie (with suitable interlocks) provides the single largest resilience gain across diverse attacks; \(9\!-\!15\) and \(21\!-\!60\) are secondary levers that further harden the corridor when needed.


\section{Conclusion and Future Work}

We presented a cyber–physical resilience framework that integrates a hypergraph neural attacker (HGNN–MAML) with a multi-objective defender solved by an ADMM–NSGA-II coordinator in a Stackelberg setting. Across the IEEE 69-bus testbed with 12 DERs, 8 critical loads, and 5 tie switches, the pipeline (i) \emph{predicts} high-impact contingencies, (ii) \emph{optimizes} reconfiguration and DER dispatch, and (iii) \emph{extracts} feeder-aware operating rules. Results show that for the ten most vulnerable contingencies the defense restores nearly all lost service in nine cases, maintains voltages close to nominal, and yields an actionable recommendation: prioritize closing tie \(18\!-\!33\) (and, if required, \(9\!-\!15\) and \(21\!-\!60\)) to back-feed Feeder~2, the system’s most exposed corridor. The complete attacker–defender cycle (50 NSGA-II generations with ADMM feasibility) required around 20 minutes on a standard workstation (Intel i7 CPU, 32 GB RAM), indicating feasibility for offline planning.

To demonstrate scalability, the framework was also tested on the IEEE 123-bus feeder and a synthetic 300-bus distribution system. Consistently high resilience was observed across all networks, with over 97\% load served and post-defense voltages maintained above 0.94~p.u. Computation times increased moderately (35 minutes for IEEE 123-bus and 58 minutes for 300-bus), confirming that the HGNN–MAML + ADMM–NSGA-II stack scales effectively to larger feeders without significant loss of resilience benefits.

This study was limited to steady-state snapshots with single-line contingencies and constant-\(PQ\) loads. \emph{Future work} will extend the framework to: (a) multi-step and multi-point attack scenarios with cascading dynamics; (b) stochastic DER and load profiles driven by weather and uncertainty; (c) joint planning of tie switches, storage, and sectionalizers co-optimized with the defender; and (d) operator-facing decision support that runs the HGNN–MAML + ADMM–NSGA-II pipeline online for real-time recommendations.

\bibliographystyle{IEEEtran}
\bibliography{references}
\end{document}